# Title: Tunable Resonator Integrated Magnetometry

Colin Stack[1], Brian Sears[1], Aruna Ramanayaka[1], David Ferguson[1], Ilya Sochnikov[2,3,4], Tony X. Zhou[1,†]

1. Northrop Grumman Mission Systems, Baltimore, MD, USA
2. Department of Physics, University of Connecticut, Storrs, Connecticut 06269, USA
3. Institute of Material Science, University of Connecticut, Storrs, Connecticut 06269, USA
4. Department of Material Science & Engineering, University of Connecticut, Storrs, Connecticut 06269, USA

†Corresponding author.

# Abstract

The quantum-technology revolution is reshaping computing, sensing, and communication. In magnetometry, recent advances leverage precise control of spin qubits and color centers in solid-state crystals for mesoscopic-scale sensing. Yet at very low temperatures, superconducting sensing technology remains unrivaled because of its non-invasiveness and higher sensitivity. Here we describe a class of superconducting sensors that offers low loss and quantum non-demolition measurement characteristics. We designed and fabricated a superconducting flux-tunable resonator (tRes) in a superconducting chip foundry and matured it to a level that combines the speed of an inductor-capacitor circuit with the flux sensitivity of a superconducting quantum interference device (SQUID) to perform magnetometry at milli-kelvin temperature to investigate targets. We introduce its fundamental functionality readily at MHz magnetic sampling rate, showcase two measurement modalities, and investigate three circuits with gradually increasing complexity to extract target-specific information. The combination of high sensitivity and fast readout characteristics make tRes an attractive and versatile magnetometer.

# Introduction

The pursuit of quantum computing has given rise to many enabling technologies due to its complexity and the associated demands on research and development. A particularly challenging requirement is high-fidelity qubit readout. In superconducting platforms, a qubit state can correspond to a minuscule persistent current, and its faithful readout can enable new sensing applications. Northrop Grumman Corporation in collaboration with MIT Lincoln lab has demonstrated a fast, lifetime-preserving readout [1] for a flux-qubit state by shuttling flux from the qubit to multiple flux-sensitive circuits and detecting it with a superconducting flux-tunable resonator (tRes) [1], [2]. In its basic form, the tRes combines a Superconducting Quantum Interference Devices (SQUID) with a superconducting resonator, thereby retaining vector field sensitivity while exploiting the fast response governed by resonator physics. Here, we expand the previous work to a more universal use of tRes as a quantum sensor.

The sensitive component in our sensor is a radio frequency (RF) SQUID. This component choice is made because SQUIDs are among the most sensitive solid-state magnetometers available. Despite serious rivalry from technologies such as NV-diamond centers [3], at low temperatures, SQUIDs remain one of the least perturbative and most sensitive magnetic-flux and magnetic-field detection technologies [4]. All SQUIDs rely on the single-valued superconducting wave function around the loop, which produces a periodic dependence on the magnetic flux threading the loop [5]. Nearly any physical property of a SQUID becomes flux-periodic, enabling its use as a flux or field sensor. SQUIDs range from large devices for bulk-material characterization, bio-magnetic signal detection, and geophysical studies [5] to sub-micron sensors [6] that have delivered outstanding signal levels and enabled discoveries in quantum materials [7], [8], [9], [10], [11], [12]. A particularly underexplored area is the use of compact SQUID sensors in scanning probe microscopy [13], [14]. These are ideally suited for demanding tasks such as diagnosing parasitic surface spins on wafer-scale materials, for instance, monolayers and especially timely, for non-perturbative superconducting circuits readout [15], [16], [17], [18], [19], [20], [21], [22]. By hybridizing SQUID and resonator to form tRes, it can potentially be suited for areas of application that remains elusive to traditional SQUIDs.

To truly become fast, non-invasive sensor, traditional SQUIDs still have a long way to go. Conventional direct current (DC) SQUIDs typically comprise two Josephson junctions embedded in a superconducting loop, which may have complex geometry [23], [24], [25], [26], [5], [27], [28], [29], [30], [31]. Under current or voltage bias, a DC SQUID enters a resistive state that introduces dissipation and back action at low temperatures, particularly in the milli-kelvin (mK) regime. Although highly sensitive, non-hysteretic DC SQUIDs are often bandwidth-limited in part by the associated electronics [5], [32], and they remain continuously biased during operation. Typical power dissipation per SQUID ranges from ~0.1 to 10 µW, depending on the critical current and shunt resistance [33]. Low-noise electronic amplification commonly employs SQUID-array amplifiers [34], which can be exceptionally quiet but, like the sensors themselves, are biased resistively and dissipate on the order of 10–100 µW. While acceptable for many 4 Kelvin applications, such dissipation is detrimental to sensing and imaging specimen at mK operation. Hysteretic DC SQUIDs may be operated in a switching mode, which can reduce overall dissipation and back action [35], nevertheless their sensitivity is generally lower than that of optimized non-hysteretic DC SQUIDs [5], limiting their utility. Another similar class of sensors, radio frequency (RF) SQUIDs—typically composed of a superconducting pickup loop and a single Josephson junction forming a resonant circuit—can, in principle, operate with less dissipation because they are not DC-biased [5]. However, conventional RF SQUIDs often exhibit reduced sensitivity due to hysteretic effects and challenges in device optimization [5], [36]. These constraints of the conventional SQUIDs motivate novel design and operation methods that enable faster, far less perturbative quantum measurements without sacrificing sensitivity. In this work, the solution is to perform magnetometry with tRes and discuss its strength in comparison to other magnetometers. In particular, we overcome the kHz dynamic bandwith challenge in SQUIDs, by operating of our tRes magnetometers below 30 mK at MHz sampling rate to extract information about target circuits.

# Main text

Here, we introduce the tRes as a dispersively read-out SQUID sensor embedded in a λ/4 microwave resonator (Fig. 1). The sensor operates via the SQUID's flux-dependent inductance and its integration within the resonator. When static or dynamic magnetic flux is applied to the SQUID loop, the critical current—and hence the loop inductance—varies periodically with flux, modulating the resonator's effective length, and thus the resonance frequency and impedance. The resulting linear or nonlinear, flux-to-frequency, and flux-to-phase transduction provides exceptional sensitivity to weak magnetic signals.

The readout component in tRes impacting its dynamic range draws inspirations from resonators which are foundational across physics. In circuits, the elementary second-order LC resonator underpins modern filters, mixers, and related components. Superconducting resonators in the GHz range have become central to microwave-frequency quantum devices [37], [38]. Introducing a Josephson junction (JJ)—a nonlinear, dissipation-less element—into a resonant circuit breaks the even level spacing of the harmonic oscillator, producing an anharmonic spectrum whose transition frequencies can be tuned by threading flux through the JJ-containing loop [39], [40], [41], [42], [43], the SQUID. The early flux-tunable resonators established key principles and design trade-offs. Today, Northrop Grumman fabricates such circuits in its own foundry [44], [45], [46] using a 13-metal-layer stack to ensure quality control, minimal crosstalk, consistent signal to noise (SNR) per measurement, a large operating parameter window, and a compact device footprint.

The designed tRes (Fig. 2d) is comprised of a quarter-wave transmission line section that is capacitively coupled to a feedline on one end, and the other end galvanically connects to an RF SQUID to ground. The tRes is designed to have flux coupled into the loop of the RF SQUID section of the resonator through its transformer input. The RF SQUID operates in the non-hysteretic regime, and its $\beta_L$ (Eq. 1) is less than one,

$$\beta_L = \frac{2\pi I_c L_0}{\Phi_0} \tag{1}$$

where $L_o = 7$ pH is the nominal geometric inductance parallel to the Josephson junction, $I_c = 35$ μA is the nominal critical current of the junction, and $\Phi_0$ is the flux quantum. Then, the RF SQUID acts as a tunable inductance as a function of flux (Eq. 2).

$$L_T(\Phi_{\text{ext}}) = \frac{L_0}{1 + \beta_L \cos\left(2\pi \frac{\Phi_{ext}}{\Phi_0}\right)} \tag{2}$$

The coupled flux modulates the inductance of the RF SQUID in the tRes resulting in a flux-dependent effective inductance of the sensor $L_T(\Phi_{ext})$. Changes in the effective inductance of the tRes inherently modifies the electrical length of tRes, and therefore shifts the resonant frequency of the tRes accordingly (Eq. 3). At the resonant frequency, the input power on the feedline is absorbed by the resonator, thus $|S_{21}|$ shown as red curve (Fig. 2a) appears as a dip at this frequency. Off resonance, the resonator is absorbing minimal power and $|S_{21}|$ is 0 dB in the ideal case.

$$f_{resonance} = \frac{1}{2\pi\sqrt{LC}} \tag{3}$$

In order to characterize the flux-tunable resonant frequency in our tRes, we use an input transformer. A flux bias line carrying a modulation current $I_{Load}$ is mutually coupled to a tRes through the input transformer coil (Fig. 2d). To readout the resonator, we use a common approach described in superconducting circuit community [1], [47], utilizing either a vector network analyzer (VNA) or a heterodyne detection with fast RF digitizers (Fig. 2b). The former suits for searching for resonances in fabricated devices and visualizing shift in resonant frequency as a current $I_{Load}$ modulates the $L_T$ (Fig. 2a). In the latter with an RF digitizer and for the remaining text, our experimental apparatus and timing sequence in Fig.2b is prepared so that the target magnetic field, or in this case, the modulation current $I_{Load}$, remains active for the duration of the RF digitizer readout. As an RF tone is transmitted, we readout the tRes on the scale of microseconds. Fig. 2c shows an arc-shaped modulation curve color-coded in blue that each $I_{Load}$ value is matched to a resonant frequency according to Eq. 3. From the equation $\Phi = M * I_{Load}$ , we convert the modulation current into flux $\Phi_{ext}$ (Fig. 2c, top axis) by multiplying our experimentally extracted, mutual inductance of 0.58 pH by the modulation current $I_{Load}$. We normalize the top axis in units of magnetic flux quantum $\Phi_0 = 2.0678 \times 10^{-15}\, Wb \cong 2\, pH * mA$ and show it in blue color corresponding the blue arcs. Since the RF SQUID inductance modulation is periodic (Eq. 2), the arc repeats every $\Phi_0$ worth in modulation current $I_{Load}$, thus resembling a stone arch bridge spanning multiple $\Phi_0$ in external flux bias $\Phi_{ext}$ (Fig. 2c). In this device design, we implement a nominal 0.5 pH mutual inductance between the flux bias line and tRes and resonator quality factor below 1000. We extract full width half maximum (FWHM) of 3.7 MHz by fitting the line cut (Fig. 2c, inset) to the resonance.

Compared to common usage of signal-to-noise (SNR), we adopt to use signal-to-background ratio (SBR) to quantify the quality of the tRes measurement (Fig. 2c):

$$SBR = \frac{I_{resonance} - I_{background}}{I_{background}} \tag{4}$$

Where symbol I defines a signal intensity. The reason to adopt SBR rather than SNR is rooted in the spectroscopic measurement where we record a physical observable as a function of frequency. In essence, when a resonant peak or a dip superimposes on top of a baseline intensity fluctuating over frequency, the baseline fluctuation cannot be simply explained by the usual noise of system or noise in the "oscilloscope" but rather the absence of spectroscopic signature of the sensor. Thereby, SBR is more suitable to use than SNR in our case and many other spectroscopy contexts. Throughout the remaining text, our signal intensity is the absolute value of scattering parameter $S_{21}$ unless we specify otherwise. In addition, our sensor manifests itself as an absorption of the energy in the RF transmission measurement, so conceptually unity SBR will constitute an ideal sensor readout. In practice (Fig. 2a and c), we observe SBR greater than 0.9, and being only slightly away from ideal showcases another desirable attribution to implementing a superconducting resonator to readout.

Compared to some other magnetometers [48], [49], within a magnetometry collected image, a magnetic contrast in background is either below 30% translating to SBR ~ 0.3 or highly sample dependent. This type of undesirably high sample dependent background may be intrinsic to the nature of measurement. For example, optics-based magnetometers experience this type of constraint when unwanted photons enter photo-detectors via optical path to increase the overall non-sensor related background intensity thereby decreasing SBR. One of the most challenging situations to use a magnetometer is when the environment or system is unstable to achieve a reasonable SBR. In that case, the sensor signal cannot be distinguished in the background, the non-sensor intensity fluctuation, and thereby its magnetometry will be considered unusable. Foreseeable scenarios for high background fluctuation may include deploying the sensor in a moving vehicle causing mechanical instability, uncontrolled temperature environment, and etc. Therefore, our magnetometer with a near unity SBR readout is highly desirable.

Our magnetometer embodies a detector signal as a function of the local sensor output signal plus intrinsic sensor noise convoluting a readout function. Next, to obtain the total measurable signal, the detector signal is amplified by a gain via a series of electronics to be multiplied by the number of readout sequences (Fig. 2b). For averaging, our magnetometry measurement has an integration time proportional to the number of repeated readout sequences. Though future work will focus on the intrinsic noise inherited at the quantum sensor, we can already see that by using superconductor and heterodyne readout method, noise in the readout function and amplification can be kept relatively minimum to achieve high SBR compared to other traditional methods. Mainly, probing at above MHz inherently helps reducing 1/f noise in the readout function, and the resonator readout method retains a series of low noise advantages by using superconducting material, Traveling-Wave Parametric Amplifier (TWPA), and cryogenic High-Electron-Mobility Transistor (HEMT) to minimize noise.

In addition to a singular transformer in tRes, a second generic flux source can also be applied simultaneously (Fig. 3a). The implementation requires a transformer network to combine both flux sources. The first transformer couples to a well-designed flux source #1, which we use as a control knob to help us gain insight into a flux source #2, unknown to us as a target. The second transformer allows us to couple flux from the target source that can be local on chip or remote at a distance emitting magnetic fields into the surrounding environment. The transformer network enables us to use the superposition of flux as a means to understand the target flux source's properties.

To demonstrate the implementation in Fig. 3a, we measure a frequency modulation curve with two flux sources present (Fig. 3b) where $I_{ext}$ equivalent to $I_{Load}$ in Fig. 2c is the flux source #1. In comparison, adding flux source #2, we engineer a static flux to cause the Fig. 3b arch bridge to shift with respect to center in x-axis. This vertical broken mirror symmetry confirms the presence of a target acting as the flux source #2. We can use the x-axis offset to quantify the amount of flux coupled in by the target. We first calibrate the on-chip mutual inductance M1 between external control flux and RF SQUID of tRes by making a line cut at a frequency below the arch bridge (Fig. 3b, inset, red trace). Finally, the offset in $I_{ext}$ away from x-axis origin is measured and converted to be $\Phi_{target} = 0.4\ \Phi_0$ static flux.

As an example of interrogating a complex target, we deploy our magnetometer to investigate a current divider circuit (Fig. 3c). Using our foremost modality, we begin by using flux dependent frequency modulation to extract circuit parameters such as mutual inductance M1. Next we develop a second modality (Fig. 3d), "iso-frequency scan", by conducting a 2-dimentional (2D) sweep of the two current sources while driving an RF tone at detuned frequency $f_0 - \delta f$, where $f_0$ is the peak frequency of the resonator at zero biased flux and $\delta f$ is the amount of frequency detuning. The detuned frequency selection determines the spacing between the intra-arc and inter-arc, where the intra-arc spacing is the current difference between resonant frequencies within the same arc color-coded in blue of Fig. 3b and 3d, and likewise inter-arc spacing is the difference between resonant frequencies in adjacent arcs. Furthermore, we utilize the slope in Fig. 3d to extract the inductances L1 and L2 in the current divider and mutual inductance M2 to be within 5% of nominal design values as expected from chip fabrication tolerance.

Next, we combine our data from the modulation curve modality and the second measurement modality, the 2D scan at an iso-frequency (Fig. 3d), to reconstruct a 3D plot in Fig. 3e, and we represent the parameters in units of flux quanta $\Phi_0$. This representation resembles the depiction in Fig. 3a, an overarching magnetometry theme using tRes as the sensor. In Fig. 3e, we map each resonant frequency to a set of flux values, $\Phi_{ext}$ and $\Phi_{target}$, derived from the measured mutual inductances M1 and M2 respectively. The arch bridge (Fig. 3e) shifts away from the $\Phi_{ext}$= 0 origin to an offset depending on the target flux ($\Phi_{target}$) also captured in demonstration of Fig. 3b. When the electric current values are converted to flux units (Fig. 3e), the slope is equal to 1 following the superposition of magnetic fields. The unit conversion simplifies the derivation of the inductances in the current divider and the mutual inductances coupled to the tRes, and ultimately upholds the superposition principle serving as the magnetometry cornerstone.

Fig. 3 demonstrates one special aspect of the tRes is tunable sensitivity in comparison to another popular quantum sensing technique based on nitrogen-vacancy-center (NV) magnetometry in diamond. By merely selecting the probing RF tone frequency, we may choose a steeper part of the frequency modulation curve (Fig. 3b, e) to achieve higher sensitivity of the magnetometer. This tunable sensitivity range enables us to tailor the magnetometry to an unknown target more readily. For instance, static spin domains and domain wall in magnetic material creates a fundamental challenge to NV magnetometry due to local magnetization gradient exhibiting strong off-axis magnetic fields away from the nitrogen-vacancy crystalline axis [50], [51]. The off-axis magnetic field is defined as the vectorial field in the plane perpendicular to NV axis and complicates the simple physical description of NV spin wavefunctions in diagonalizable basis, and this results in degraded spin dependent optical contrast of NV magnetometry. Furthermore, NV electronic spin resonance (ESR) is not linearly dependent to off-axis field to cause additional inaccuracy in magnetometry [50]. In contrast to using tRes magnetometry, this type of undesirably strong field amplitude can be optimally tailored by tunable sensitivity selection, for example, our static flux-modulation method by $\Phi_{ext}$ on-chip (Fig. 3b, e) to achieve highly efficient sensor detection.

As a comprehensive example of the application, we deploy the tRes to target a superconducting flux memory circuit (Fig. 4a) to extract its on-chip parameters. We design the flux memory device to encode memory within its inductive load in steps of quantized flux. This type of device also operates as a DC current source for flux-biasing another superconducting circuit element. The flux bias stored in memory remains in the form of a persistent current inside the element even when all bench-top current sources are powered off. Flux memory does not dissipate on-chip passive power and is ideal for eliminating the number of active DC current sources on the equipment rack.

In the experiment (Fig. 4c) to test flux memory, we adopt the iso-frequency scan measurement modality and perform two delta current bias, $I_{\Delta}$, sweeps in opposite directions as one hysteretic operation loop for the memory cell, and simultaneously measure our tRes magnetometer which is non-hysteretic. Due to the hysteretic nature of the memory cell, we sweep $I_{ext}$ first in vertical line trace entirely before stepping $I_{\Delta}$ horizontally. After $I_{\Delta}$ reaches 0.5 mA, $I_{\Delta}$ is stepped back down to -0.5 mA to complete the full hysteretic loop. While $I_{\Delta}$ and $I_{ext}$ are held constant at each pixel (Fig.2b), we record an $S_{21}$ measurement at a detuned tRes frequency. A starting $I_{\Delta}$ current bias ramp is programmed before the experiment to set the memory to an initial state.

For simulation (Fig. 4b) and modeling to understand our data, we use WRspice to build a test bench based on the circuit in Fig. 4a and implement our nominal design parameters for the flux memory and tRes. Assuming no junction asymmetry, when the $I_{\Delta}$ magnitude of the memory cell exceeds $2I_c$, the junctions allow for a phase slip and inject or expel $\Phi_0$ into the $\Delta$ loop depending on the bias polarity and its current state. This happens repeatedly as the $I_{\Delta}$ magnitude increases. This test bench operation produces the steep sloped ladders in steps of single flux quanta (SFQ) that appears in Fig. 4b, c. The slope of these ladders is equal to ratio of the mutual inductances M2 and M1. We extract the M2 mutual inductance to be 12.986 pH which is within 4% of the nominal design value of 12.5pH. Two intriguing flatter sections appears in each Fig. 4b, c, hinting that resonant frequency remains relatively unchanged for different $I_{\Delta}$. Our investigation reveals that the $I_{\Delta}$ bias polarity has changed at the onset of flatter sections. When the $I_{\Delta}$ polarity changes, a screening current is flowing through the M2 load to produce the shallower slope until $I_{\Delta}$ is great enough to trigger the memory cell's junctions for climbing the SFQ-stepped ladder. Our experimental data and WRspice simulation show strong agreement and further showcase the versatility of the iso-frequency scan measurement modality.

The mutual inductance, $M$, between the tRes and the magnetic sensing target constitutes the central calibration parameter of our approach. Once $M$ is extracted, the tRes response can be quantitatively related to the intrinsic magnetic susceptibility of the system, rather than remaining a geometry-dependent frequency shift. This generalizes the measurement principle beyond the specifics of the circuit and allows one to regard $M$ as a universal coupling factor between a microwave probe and an arbitrary magnetic object.

The role of $M$ in this context is directly analogous to its function in conventional SQUID-based AC susceptibility measurements, where the SQUID detects changes in mutual inductance between a drive coil and a pickup loop mediated by the sample's response. In that framework, variations in $\chi(\omega)$ alter the detected flux, which the SQUID transduces into measurable in-phase and out-of-phase signals [14]. The tRes implements a similar principle in the microwave regime, thereby extending the technique to higher frequencies with dispersive readout and operation at mK temperature range where novel quantum physics resides.

We envision that with knowledge of $M$, the tRes could become a quantitative tool for low-drive, high-bandwidth spectroscopy of quantum materials. This capability enables, for example, studies of vortex dynamics [52], [53], pair-breaking processes [54], [55], and kinetic-inductance nonlinearities in superconductors [56], magnon resonances [48], and domain-wall dynamics in two-dimensional magnets and van der Waals heterostructures [57]; slow relaxation phenomena in frustrated or glassy spin systems [58], and defect-based ensembles that couple to microwave magnetic fields and set loss tangents in quantum circuits [59]. In each of these cases, extracting the frequency-dependent susceptibility, $\chi(\omega)$, via $M$ provides direct access to intrinsic material parameters—such as relaxation times, damping constants, or pinning strengths—rather than to device-specific observables.

More broadly, treating $M$ as a central, geometry-independent characteristic elevates the tRes from a circuit-specific readout device to a general framework for quantum sensing. Accurate calibration of $M(\omega)$ enables the conversion of raw frequency or phase shifts into intrinsic $\chi(\omega)$, granting access to both dispersive ($\chi'$) and dissipative ($\chi''$) channels with high spectral resolution. This approach provides a pathway from static susceptometry toward full GHz-range spectroscopy of magnetic excitations, including magnon resonances, vortex depinning, and defect spin relaxations, thereby establishing tRes as a versatile probe of diverse quantum materials.

In summary, we have demonstrated magnetometry by using a tRes for quantum sensing applications at millikelvin temperature range. The sensors exhibit low noise, high SBR quality which is important for future sensing applications, such as scanning probe microscopy imaging specimen inside dilutional fridge. In particular, the tradeoff of SBR is integration time where raster image capture needs optimization to avoid overly long measurement time. Our high SBR will withstand the rapid magnetic image generation required by microscopy. Furthermore, we have demonstrated different measurement modalities that allow for efficient flux modulation, optimal operation point adjustments, and extraction of critical parameters of a tested structure such as the mutual inductance between the sensor and the target device. Future work includes expansion of the operation range to faster measurements and optimization of the sensor geometry for scanning probe imaging and remote off-chip sensing applications.

## Acknowledgment

The Northrop Grumman team would like to acknowledge a support by from the Army Research Office via Contract No. W911NF-25-C-0007. I. S. would like to acknowledge a support by from the Army Research Office via Contract No. W911NF-24-C-0093. I.S. would also like to acknowledge a partial support from UCONN QuTech program. We thank R. Burnett for helpful discussions and M. Savoury for graphic design. Any opinions, findings and conclusions or recommendations expressed in this material are those of the author(s) and do not necessarily reflect the views of the Army Research Office and the Army Contracting Command. All trademarks are the property of their respective owners. Approved for Public Release: Distribution Unlimited; NG26-0881.

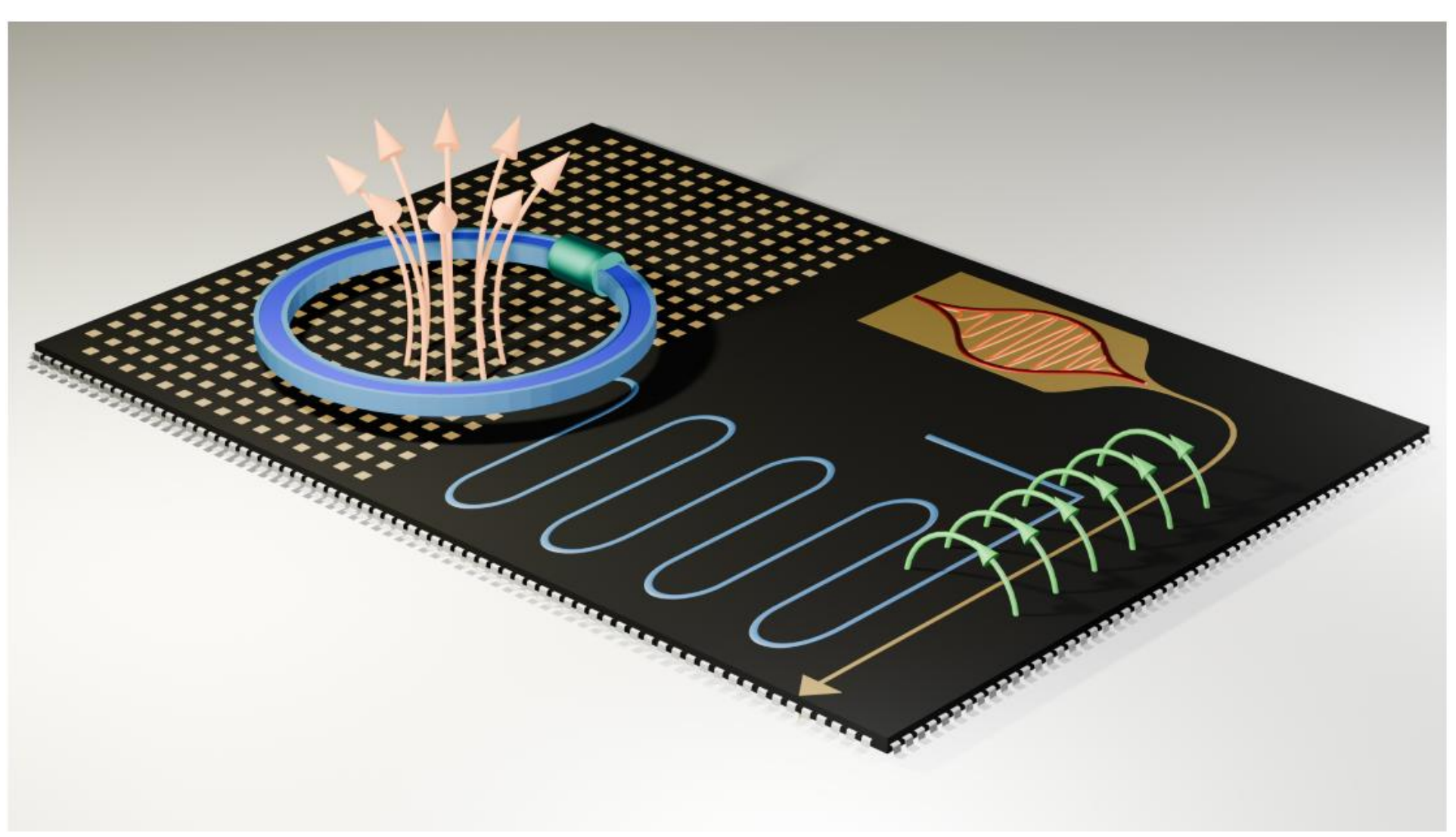

Fig. 1. A tunable resonator operating as a magnetometer to sense flux from targets emitting magnetic fields into their surrounding space. Graphic credit: Matthew Savoury.

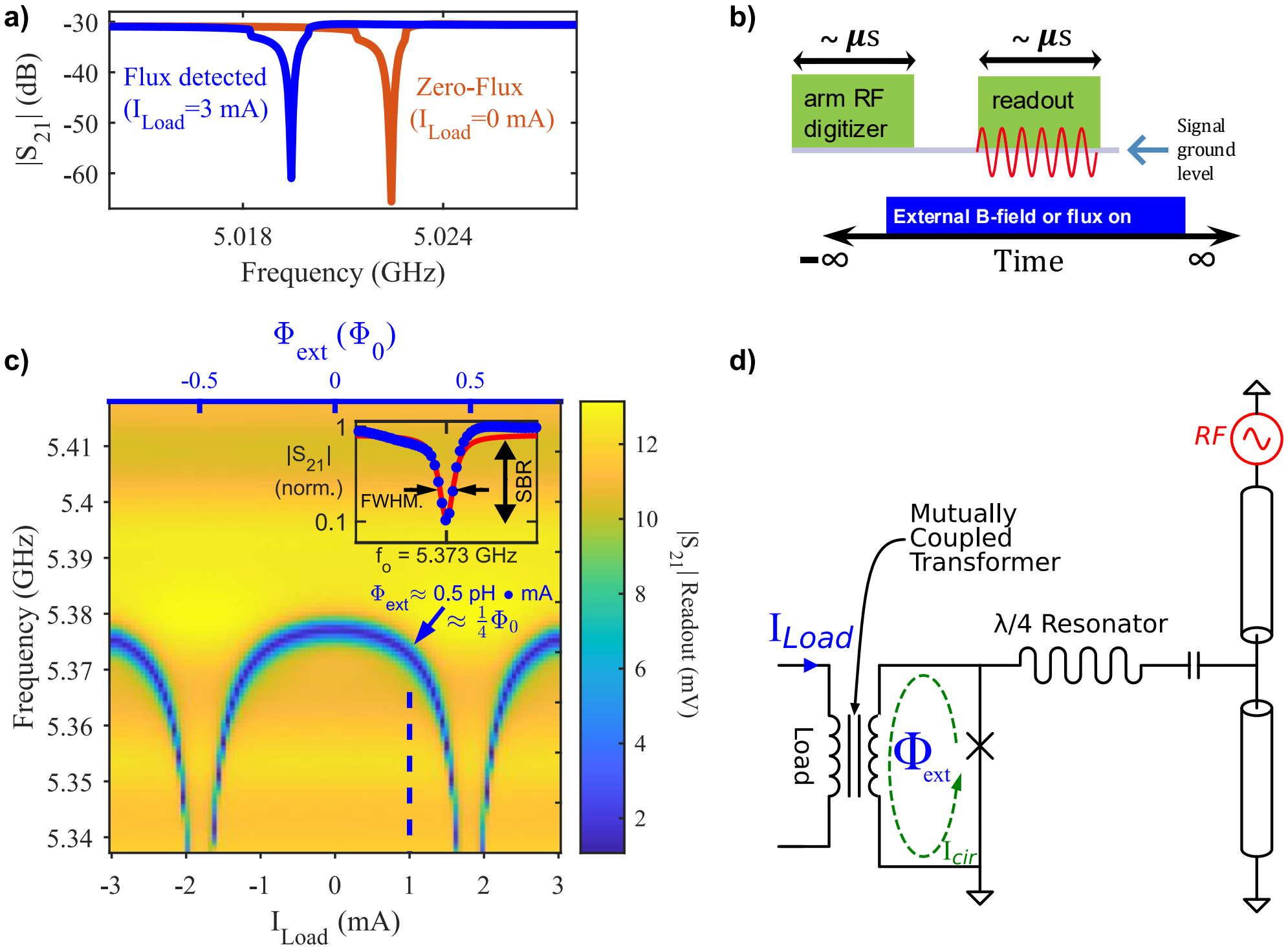

Fig. 2. Frequency modulation of a tRes. (**a**) An $S_{21}$ measurement with 3 mA load current on and off (blue, orange, respectively). This demonstrates tRes detection of an electrical current. (**b**) The readout sequence in a tRes magnetometry experiment. (**c**) Frequency modulation curve that maps each modulation current value $I_{Load}$ (bottom axis) or external flux $\Phi_{ext}$ (top, blue axis) to a resonant frequency. (Inset) We made a vertical line-cut at $I_{Load}$ of 1 mA (dash line) to examine $|S_{21}|$, and we implement a Lorenzian fit function to describe the response. The fitting function extracted a FWHM of 3.7 MHz centered at $f_o$ equal to 5.373 GHz, and this characterization of the tRes demonstrates a near-unity SBR ratio. (**d**) A current $I_{Load}$ (blue) generates an external flux $\Phi_{ext}$ (blue) threading through RF SQUID of the tRes which is read out in the GHz frequency range via an RF generator (red). $\Phi_{ext}$ induces a circulating current (green) and modulates the phase across the junction to shift $L_T$.

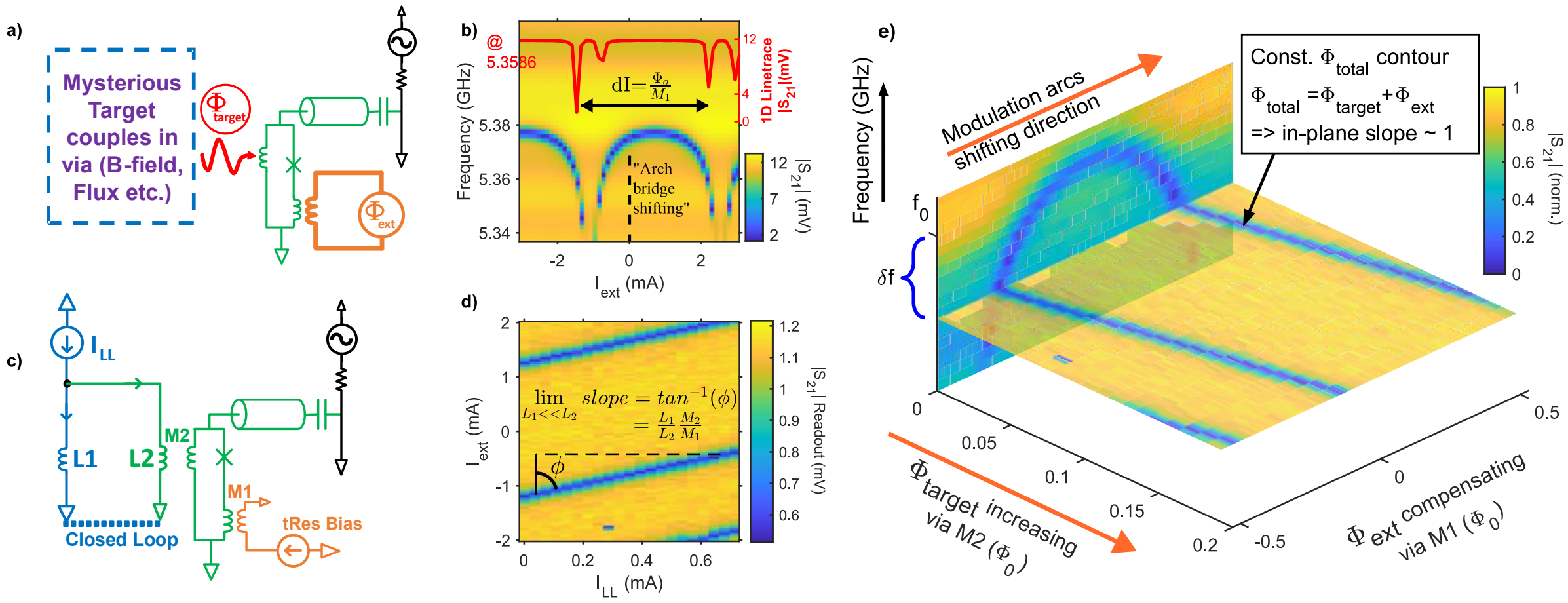

Fig. 3. tRes magnetometer deployment for quantum sensing. (**a**) Target magnetic fields or flux (red) are sensed by a tRes, and an external flux control (orange) assists to investigate the unknown traits of the target. (**b**) We fabricated a circuit to illustrate the general concept in (a). A frequency modulation curve by current in the circuit, not shown here, is measured. A flux introduced from the target shifts the center of the modulation curve away from the origin by an offset in x-axis, so the stone arch bridge suspends asymmetrically over the x-axis. Dash line guides the eye to indicate the vertical mirror line at $I_{ext} = 0\ mA$, the center of x-axis. (Inset) A 1D line-trace (red) is made at 5.359 GHz. Due to the periodic nature of the tRes every $\Phi_0$, the difference in $I_{ext}$ between the resonant peaks divided by $\Phi_0$ yields the mutual coupling from the tRes to the $I_{ext}$ source. (**c**) A bias line with current $I_{ext}$ and a current divider where L1 << L2 are mutually coupled to a tRes (green). (**d**) A 2D sweep of $I_{ext}$ and current divider bias $I_{LL}$ using the iso-frequency scan measurement modality. The slope is equal to the ratio of the mutual inductances multiplied by the ratio of the inductors in the current branch circuit being sensed. (**e**) Merging the tRes magnetometer measurement modalities. The electrical current from $I_{LL}$ and $I_{ext}$ are converted to quantized flux coupling to the tRes, $\Phi_{ext}$ and $\Phi_{target}$ respectively. (Inset) As a result of the superposition of the magnetic fields in the tRes, the in-plane slope is equal to one to maintain a constant total flux in the tRes.

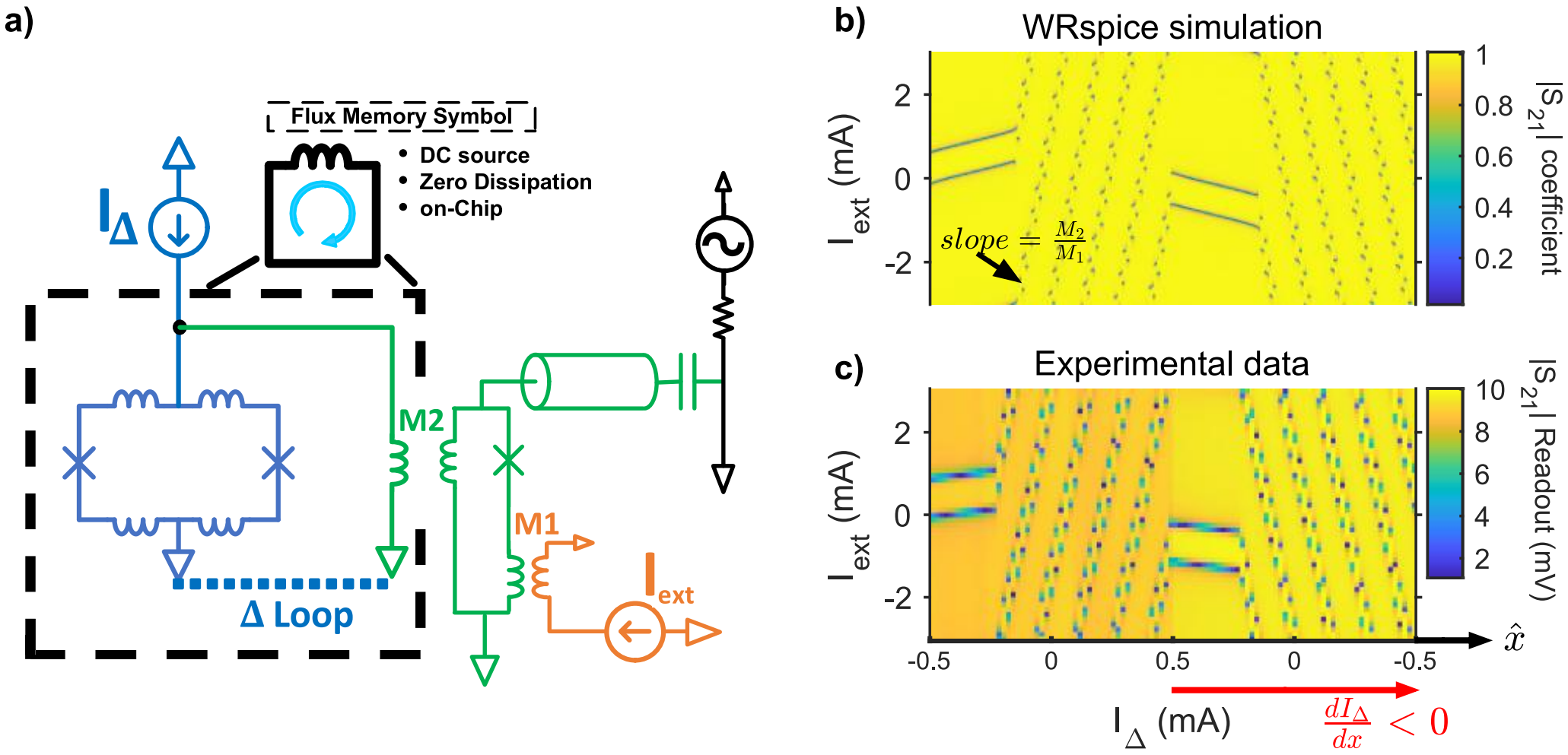

Fig. 4. Performing tRes magnetometry on a superconducting flux memory device. (**a**) A tRes (green) is mutually coupled to a flux memory cell (inside dash line) with a current bias $I_\Delta$ through the DC SQUID (blue in dash lined box), and a external current $I_{ext}$ flux biases the tRes as an independent control knob (orange). (**b**) Simulation of the circuit in (a) to show one to one model for the iso-frequency measurement modality. The slope of the steep ladder is the ratio of the mutual inductance from the memory cell to tRes over the mutual inductance of the $I_{ext}$ bias to the tRes. The simulation results validate the experimental data and demonstrate the effectiveness of iso-frequency modality to extract device parameters and gain insight into our target. (**c**) Experimental data from the circuit in (a). The $\hat{x}$-axis consists of a complete hysteretic loop in $I_\Delta$ sweep with first half of the loop with increasingly positive in step and second half of the loop shown in red arrow stepping values increasingly negative (red).